\documentclass[journal,twoside]{IEEEtran}

\usepackage{amsmath}
\usepackage{latexsym, amssymb}
\usepackage{graphicx}
\usepackage{amsbsy}
\usepackage{amsopn, amstext}
\usepackage{hyperref}
\usepackage{cancel, color}
\usepackage{epstopdf}
\usepackage{float}
\usepackage{subfigure}
\usepackage{cite}
\usepackage{stfloats}
\usepackage{algorithmic}
\usepackage{setspace}
\usepackage[ruled]{algorithm2e}

\usepackage{booktabs}
\newcommand{\tabincell}[2]{\begin{tabular}{@{}#1@{}}#2\end{tabular}}

\newtheorem{theorem}{Theorem}
\newtheorem{lemma}{Lemma}

\newtheorem{example}{Example}
\newtheorem{remark}{Remark}

\newtheorem{assumption}{Assumption}

\hypersetup{hypertex=true,
            colorlinks=true,
            linkcolor=black,
            anchorcolor=black,
            citecolor=black}

\begin{document}

\title{Global Regulation of Feedforward Nonlinear Systems: A Logic-Based Switching Gain Approach}

\author{Debao Fan, Xianfu Zhang$^*$, \IEEEmembership{Member, IEEE}, Gang Feng, \IEEEmembership{Fellow, IEEE}, and Hanfeng Li

\thanks{Debao Fan, Xianfu Zhang, and Hanfeng Li are with the School of Control Science and Engineering, Shandong University, Jinan 250061, P.R. China (e-mail: fandebao@sdu.edu.cn; zhangxianfu@sdu.edu.cn; lihanfeng@sdu.edu.cn).}

\thanks{Gang Feng is with the Department of Biomedical Engineering, City University of Hong Kong, Hong Kong (e-mail: megfeng@cityu.edu.hk).}
}

\maketitle

\begin{abstract}
In this article, we investigate the global regulation problem for a class of feedforward nonlinear systems. Notably, the systems under consideration allow unknown input-output-dependent nonlinear growth rates, which has not been considered in existing works. A novel logic-based switching (LBS) gain approach is proposed to counteract system uncertainties and nonlinearities. Furthermore, a \emph{tanh}-type speed-regulation function is embedded into the switching mechanism for the first time to improve the convergence speed and transient performance. Then, a switching adaptive output feedback (SAOF) controller is proposed based on the developed switching mechanism, which is of a concise form and low-complexity characteristic. It is shown that the objective of global regulation is achieved with faster convergence speed and better transient performance under the proposed controller. Moreover, by strengthening the switching mechanism, the improved control approach can deal with feedforward nonlinear systems with external disturbances. Finally, representative examples are presented to demonstrate the effectiveness and advantages of our approach in comparison with the existing approaches.
\end{abstract}

\begin{IEEEkeywords}
Feedforward nonlinear systems, global regulation, output feedback, logic-based switching gain.
\end{IEEEkeywords}

\IEEEpeerreviewmaketitle

\section{Introduction} \label{lintro}

\IEEEPARstart{F}{eedforward} nonlinear systems, which are also called upper-triangular nonlinear systems, are an important class of nonlinear systems \cite{KM2004Feedback,KI2012Global,ZJ2020Local,LK2023Dlc}. Many practical systems can be transformed into feedforward nonlinear systems by appropriate transformations, e.g., the cart-pendulum system \cite{ZCR2014Global} and the inertia wheel pendulum system \cite{ZB2018Global}.

As is well known, global regulation of feedforward nonlinear systems is a fundamental and challenging topic \cite{KMS2012Global}. The objective of global regulation is to design an appropriate controller to guarantee the boundedness of all signals of the resulting closed-loop system and the convergence of the system states for any initial conditions \cite{LHF2022An}. In recent years, the global regulation problem has been widely studied, see  \cite{KMS2012Global,LHF2022An,JXL2017Global,YX2009Pseudo,KM2010Input,DSH2010Global}. The control approaches proposed therein are roughly divided into two categories: \emph{the gain scaling approach} \cite{KMS2012Global,LHF2022An,JXL2017Global}, and \emph{the forwarding design approach} \cite{YX2009Pseudo,KM2010Input,DSH2010Global}. Similar to the well-known backstepping design approach in \cite{KM1995Global,LY2023Btd,WH2017Ofa,SS2020Gfc,LY2023Efc,SZY2022Aef}, the forwarding design approach is an iterative design strategy that often leads to a very complex controller. Moreover, as the system order increases, the design procedures become more tedious. In contrast to the forwarding design approach, the gain scaling approach is not an iterative design strategy which often leads to a much simpler controller \cite{ZXF2011Feedback}. In fact, the gain scaling approach is even more appealing in the framework of output-feedback control \cite{ZXF2011Design}.

For the gain scaling approach, four types of gains are typically used and they are \emph{static}/\emph{constant gain}, \emph{time-varying gain}, \emph{dynamic gain}, and \emph{LBS gain}. When the system uncertainty or nonlinearity is intrinsically minor, a static/constant gain (see, e.g., \cite{LHF2021Leader-follower,ZCH2014Leader-follower}) or a bounded time-varying gain (see, e.g., \cite{LHF2021Leader-follower}) is sufficient to meet the desired requirement. Otherwise, when the uncertainty or nonlinearity is more significant, an unbounded time-varying gain (see, e.g., \cite{LQR2020Global}) or a dynamic gain (see, e.g., \cite{LHF2022An, ZXF2011Design, LZS2015Design}) is often needed to suppress their negative effects. However, first, as mentioned in \cite{LH2006Universal}, it is impractical to apply unbounded time-varying gains to achieve control objectives. Second, the dynamic gains typically require continuous and considerable gain updates, which contributes to the high-complexity characteristics of the resulting controllers. Moreover, the dynamic gains perform poorly when dealing with serious uncertainties and nonlinearities with  unknown input-output-dependent growth rates \cite{WP2019Global, GC2020Global}. In particular, as shown in \cite{WP2019Global}, a single dynamic gain can usually only handle the case of $p\in [ 0,\frac{1}{2n}) $, where $p$ is the power of the system output. Although the restriction on $p$ is relaxed in the case of constructing a dual dynamic gain, the complexity of the resulting controller is greatly increased and the type of growth rates is restricted to the special polynomial function of control input \cite{GC2020Global}. Furthermore, most of these mentioned gain scaling approaches suffer from slow convergence speed and poor transient performance \cite{LHF2022An,ZXF2011Feedback,ZXF2011Design,LHF2021Leader-follower,ZCH2014Leader-follower}. It is clear that the more significant the uncertainties and nonlinearities, the more difficult their suppression. Then, the following questions naturally arise: (i) \emph{For feedforward nonlinear systems, is it possible to tolerate more general nonlinear growth rates?} (ii) \emph{What kind of control approaches can be developed to handle uncertainties and nonlinearities with such nonlinear growth rates?}

It is well known that the LBS control is highly effective in compensating for a variety of uncertainties and nonlinearities, see \cite{YX1999Global, VL2011Supervisory, YM2020Logic-based, HC2021Output, CW2018Global, HC2022Pos, HY2019Switching}. However, no work has been done to apply a LBS gain approach to address the global regulation problem of feedforward nonlinear systems with unknown input-output-dependent growth rates. This motivates the research of this article: \emph{is it possible to design a novel LBS gain approach to counteract more serious uncertainties and nonlinearities while improving control performance?} For this, a novel LBS mechanism is developed, where the logic unit supervises the available state information of a properly-defined Lyapunov function. After finite switching actions, global regulation can be finally achieved and satisfactory control performance can be guaranteed. The contributions of this article are summarized as follows:

\begin{itemize}
\item \emph{\textbf{In terms of system model}}: The considered system model is more general in comparison to \cite{LHF2022An,ZXF2011Design,LQR2020Global,LZS2015Design,WP2019Global} in terms of the following two aspects: (i) the nonlinear growth rates can be a combination of unknown constant, a polynomial function of system output with maximum power $p\in [ 0,\frac{1}{n}) $ and an arbitrary function (rather than polynomial function) of control input; and (ii) external disturbances are allowed to be unknown and time-varying. More detailed comparisons can be found in Table \ref{table-upper-comparisons}. In fact, global regulation of feedforward systems with such nonlinear growth rates and external disturbances is still an open problem.
\end{itemize}

\begin{itemize}
\item \emph{\textbf{In terms of control approach}}: Different from the existing traditional control approaches, e.g., the forwarding design approach in \cite{YX2009Pseudo,KM2010Input,DSH2010Global}, the static/constant gain approach in \cite{LHF2021Leader-follower,ZCH2014Leader-follower}, the time-varying gain approach in \cite{LHF2021Leader-follower,LQR2020Global}, and the dynamic gain approach in \cite{LHF2022An,ZXF2011Design}, a novel LBS gain approach is proposed to address the global regulation problem of feedforward nonlinear systems. The SAOF controller designed by this approach is low-complexity. Moreover, the framework established by this approach can provide an effective tool for relevant problems of other systems.
\end{itemize}

\begin{itemize}
\item \emph{\textbf{In terms of control performance}}: A novel \emph{tanh}-type speed-regulating function is embedded in the switching mechanism, which is beneficial to improve the convergence speed and transient performance of the resulting closed-loop system. This work contrasts with many existing works, such as \cite{ZXF2011Design,ZXF2011Feedback,ZCH2014Leader-follower}, where the issue of control performance is not addressed or the performance is quite poor.
\end{itemize}

The remainder of this article is organized as follows. Section \ref{Ppf} describes the preliminaries and problem formulation. A SAOF control strategy is designed in Section \ref{Sec3}, and the disturbance tolerance is discussed in Section \ref{Sec4}. Section \ref{Examples} presents examples and comparisons. In the end, Section \ref{Conclusion} gives a conclusion.

\section{Preliminaries and Problem Formulation}  \label{Ppf}

\subsection{LBS Gain}

For ease of understanding, the basic concept of LBS gain is introduced. As the name implies, the gain is adjusted online by a switching mechanism, and presented in a piecewise constant manner. For this, we use $r_{\ell(t)}\geq 1$ to denote the piecewise constant gain, where $\ell(t)$ is a switching signal. The switching moments are denoted as $t_{m}$, $m =1,2,\ldots$, with the initial moment $t_{1}=0$. For $t\in [t_{m},t_{m+1}) $, $m =1,2,\ldots$, there holds $\ell(t)=m$. In what follows, we will describe the dynamical systems in an arbitrary switching interval $[t_{m},t_{m+1}) $, where $m =1,2, \ldots$.

\begin{figure}[tbp]
\par
\begin{center}
\includegraphics[height=4.7cm, width=3.5in]{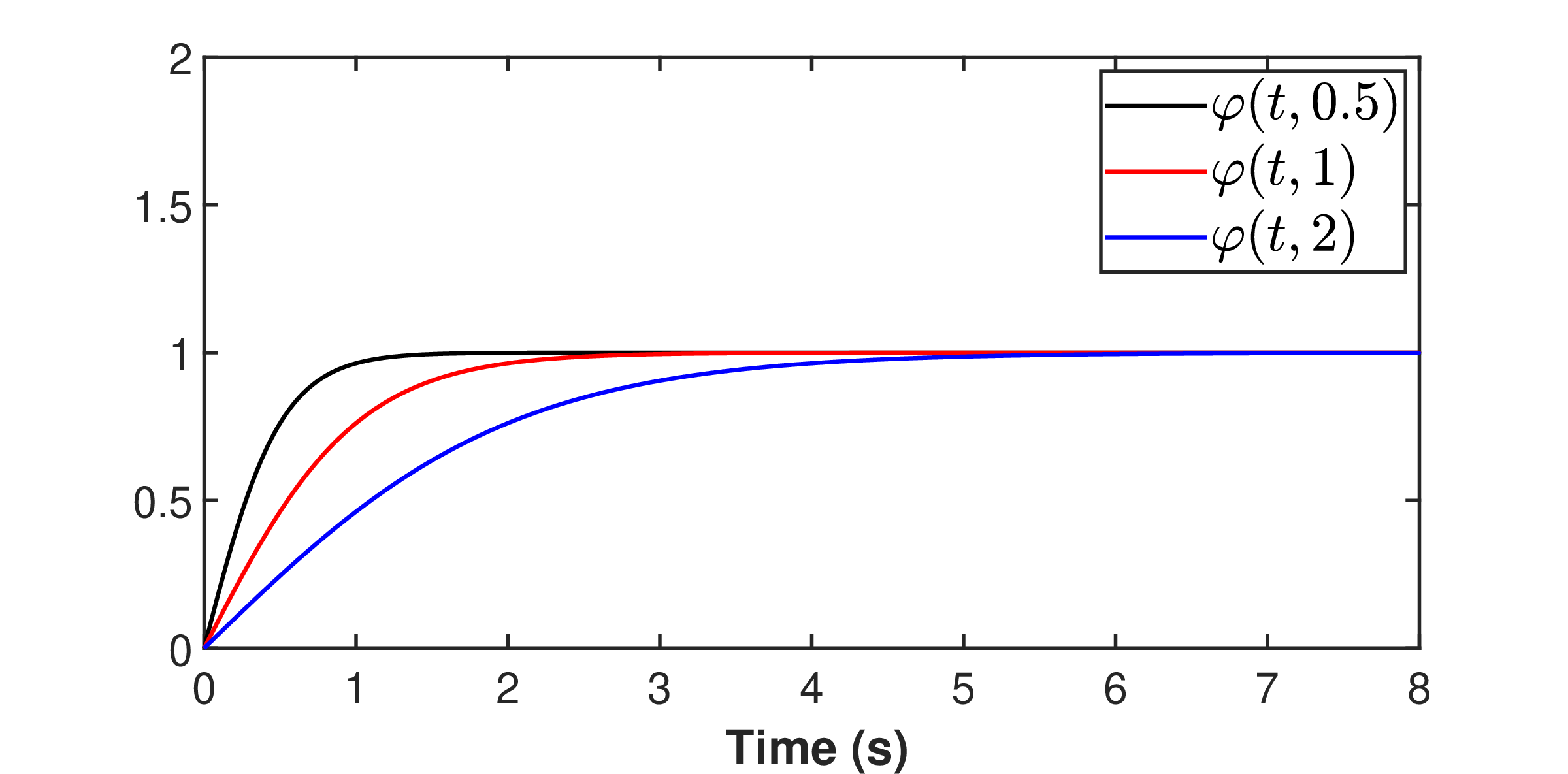}
\end{center}
\par
\caption{The trajectories of speed-regulating function $\varphi (t,\mu )$ under different adjustable constants.}
\label{Fig-S-r-f}
\end{figure}

\subsection{Speed-Regulating Function}

To improve the control performance, a novel \emph{tanh}-type speed-regulating function $\varphi(t,\mu)$ is constructed as
\begin{eqnarray} \label{srf1}
\varphi (t,\mu )=\tanh \Big( \frac{t}{\mu }\Big),~~ \text{for}  ~ t\in \left[ 0,+\infty\right),
\end{eqnarray}
where $\tanh \big( \frac{t}{\mu }\big) =\frac{2}{1+e^{-\frac{2t}{\mu }}}-1$, and $\mu >0$ is an adjustable bounded constant.

\begin{lemma}  \label{srf-lemma}
The speed-regulating function $\varphi (t,\mu )$ possesses the following unique properties:
\begin{itemize}
\item $\varphi (t,\mu )$ is strictly increasing for $t\in \left[0,+\infty \right) $ with $\varphi (0,\mu )=0$ and $\lim\limits_{t\rightarrow +\infty } \varphi(t,\mu )=1$;
\item $t \cdot \varphi (t,\mu )\geq t-0.2785\mu, ~~ \text{for}  ~ t\in \left[ 0,+\infty\right).$
\end{itemize}
\end{lemma}

\begin{IEEEproof}
Note that $\varphi (t,\mu)$ is well defined for $t\in \left[0,+\infty \right) $. From \eqref{srf1}, we get
\begin{eqnarray} \label{srf2}
\dot{\varphi}(t,\mu )=\frac{4}{\mu e^{\frac{2t}{\mu }}\big( 1+e^{-\frac{2t}{\mu }}\big) ^{2}}, ~~ \text{for}  ~ t\in \left[ 0,+\infty\right).
\end{eqnarray}

By \eqref{srf2}, one can straightforwardly obtain that $\dot{\varphi}(t,\mu )>0$ for $t\in \left[ 0,+\infty \right) $. Thus, $\varphi(t,\mu )$ is strictly increasing for $t\in \left[ 0,+\infty \right) $. It then follows that $\varphi (t,\mu )$ has its minimum value of $0$ at $t=0$ and converges to its maximum value of $1$ as $t\rightarrow +\infty $. Moreover, through the simple transformation of inequality (24) in \cite{XL2017Adaptive}, we can easily verify the second property.
\end{IEEEproof}

\begin{remark} \rm  \label{srf-remark1}
The hyperbolic tangent function has been widely used in artificial neural networks over the past few decades. Unlike the existing works, we propose to use it to regulate the switching speed. Specifically, as shown in Fig. \ref{Fig-S-r-f}, the growth speed of $\varphi $ is affected by the value of $\mu $, i.e., the larger $\mu $ is, the slower $\varphi $ grows. This function will be embedded into our switching mechanism to be shown in the next section so that the control objective with fast convergence speed and desirable transient performance can be achieved.
\end{remark}

\subsection{Problem Formulation}
Consider the following feedforward nonlinear system:
\begin{eqnarray} \label{pf1}
\left\{
\begin{array}{l}
\dot{x}_{i}=x_{i+1}+f_{i}(t,x,u), \\
\dot{x}_{n}=u, ~~ i=1,2,\ldots ,n-1, \\
y=x_{1},
\end{array}
\right.
\end{eqnarray}
where $x=\left[ x_{1},x_{2},\ldots ,x_{n}\right] ^{\mathrm{T}}\in \mathbb{R}^{n}$, $u\in \mathbb{R}$ and $y\in \mathbb{R}$ are the system state, control input and system output, respectively; unknown nonlinear functions $f_{i}(\cdot )$: $\mathbb{R}^{\geq0}\times\mathbb{R}^{n}\times\mathbb{R}\rightarrow\mathbb{R}$, $i=1,2,\ldots ,n-1$, are piecewise continuous in the first argument and continuous in the other arguments. Next, an assumption is imposed on system \eqref{pf1}.

\begin{assumption} \rm \label{a2}
For $i=1,2,\ldots ,n-1$, there exist a known nonnegative continuous function $\gamma (u)$, a known nonnegative constant $p$, and an unknown nonnegative constant $\theta$, such that
\begin{eqnarray} \label{pfa2}
\begin{aligned}
\vert f_{i}(\cdot )\vert \leq  \theta \gamma ( u) (1+\vert y\vert ^{p}) \bigg(\sum_{j=i+2}^{n+1} \vert x_{j}\vert +\vert u\vert \bigg),
\end{aligned}
\end{eqnarray}
where $p\in [ 0,\frac{1}{n}) $ and $x_{n+1}=0$.
\end{assumption}

\begin{remark} \rm  \label{srf-remark-XJ1}
Generally, the system \eqref{pf1} under Assumption \ref{a2} is called a feedforward system with an unknown input-output-dependent growth rate. It is worth noting that Assumption \ref{a2} in this article is more general than the common assumptions for feedforward nonlinear systems, largely owing to the following two factors. First, the unknown constant $\theta$ allows significant uncertainties for the system. Second, Assumption \ref{a2} indicates that the system allows inherent nonlinearities described by $\gamma ( u) (1+\vert y\vert ^{p})$. The serious uncertainties and nonlinearities make Assumption \ref{a2} applicable to most feedforward nonlinear systems. Table \ref{table-upper-comparisons} presents more comparisons between our work and several relevant works.
\end{remark}

\section{Switching Adaptive Output Feedback Control} \label{Sec3}

In this section, a SAOF control strategy is proposed for system \eqref{pf1} under Assumption \ref{a2}.

\subsection{Design of SAOF Controller} \label{Sec3.1}

Inspired by \cite{LHF2022An} and \cite{ZXF2011Design}, a novel low-gain observer is designed as follows, for $t\in \left[ t_{m},t_{m+1}\right) $,
\begin{eqnarray} \label{ff-1.1}
\left\{
\begin{array}{l}
\dot{\hat{x}}_{i}=\hat{x}_{i+1}+a_{i}r_{m}^{-i}( y-\hat{x}_{1}),~~ i=1,2,\ldots ,n-1, \\
\dot{\hat{x}}_{n}=u+a_{n}r_{m}^{-n}( y-\hat{x}_{1}),
\end{array}
\right.
\end{eqnarray}
where $a_{i}$, $i=1,2,\ldots ,n$, are the coefficients of a Hurwitz polynomial $h _{1}(s)=s^{n}+a_{1}s^{n-1}+\cdots +a_{n-1}s+a_{n}$, and $r_{m}$ is a LBS gain.

\begin{table}[t]
	\caption{Comparisons between our work and existing relevant works.}		
    \label{table-upper-comparisons}
	\centering
    \renewcommand\arraystretch{1.06}
	\setlength{\tabcolsep}{1.56mm}{
	\begin{tabular}{lllll}
		\toprule[1pt]
		Reference & \tabincell{l}{Growth  rate \\ ($p$, $\theta$, $\gamma(u)$) } &   \tabincell{l}{Disturbance  \\ evaluation} & \tabincell{l}{Control \\ approach} \\
		\midrule[0.5pt]
		\cite{LHF2022An} &  \tabincell{l}{$p=0$ } &  \tabincell{l}{No, \\ i.e., $\varpi_i=0$ } & \tabincell{l}{Dynamic \\ gain}\\
		\midrule[0.5pt]
		\cite{ZXF2011Design} & \tabincell{l}{$p=0$; $\theta $ is a \\ known constant} &  \tabincell{l}{No, \\ i.e., $\varpi_i=0$ } & \tabincell{l}{Dynamic \\ gain}\\
		\midrule[0.5pt]
		\cite{LQR2020Global} & \tabincell{l}{$p=0$;  $\gamma ( u)$ \\ is a constant } &  \tabincell{l}{No, \\ i.e., $\varpi_i=0$ }  & \tabincell{l}{Time-varying \\ gain} \\
		\midrule[0.5pt]
		\cite{LZS2015Design}	& \tabincell{l}{$p=0$;  $\theta$ is a \\ known constant} & \tabincell{l}{No, \\ i.e., $\varpi_i=0$ } & \tabincell{l}{Dynamic \\ gain }\\
		\midrule[0.5pt]
		\cite{WP2019Global}	&  \tabincell{l}{$p\in [ 0,\frac{1}{2n}) $  }  &  \tabincell{l}{No, \\ i.e., $\varpi_i=0$ } & \tabincell{l}{Dynamic \\ gain} \\
		\midrule[0.5pt]
		\tabincell{l}{This \\ article} &  \tabincell{l}{$p\in [ 0,\frac{1}{n})$; $\theta$ is \\ an unknown \\ constant; $\gamma ( u)$ \\ is a function }  &  \tabincell{l}{Yes, \\ i.e., $\varpi_i\neq0$ } & \tabincell{l}{LBS gain} \\
		\bottomrule[1pt]
	\end{tabular}}
\end{table}

Define the following state transformation, for $t\in\left[t_{m},t_{m+1}\right) $,
\begin{eqnarray} \label{ff-1.2}
\eta _{i}=\frac{\hat{x}_{i}}{r_{m}^{n-i+1}},~~ \varepsilon _{i}=\frac{x_{i}-\hat{x}_{i}}{r_{m}^{n-i+1}},~~ i=1,2,\ldots ,n.
\end{eqnarray}

The SAOF controller is designed as follows:
\begin{eqnarray} \label{ff-1.3}
u=-\sum_{i=1}^{n}b_{i}\eta _{i},~~ \text{for}  ~ t\in\left[ t_{m},t_{m+1}\right),
\end{eqnarray}
where $b_{i}$, $i=1,2,\ldots ,n,$ are the coefficients of another Hurwitz polynomial $h _{2}(s)=s^{n}+b_{n}s^{n-1}+\cdots +b_{2}s+b_{1}$.

Then, for $t\in\left[t_{m},t_{m+1}\right) $, by \eqref{ff-1.1}-\eqref{ff-1.3}, we have
\begin{eqnarray} \label{xjff-3.1.1}
\left\{
\begin{array}{l}
\dot{\eta}_{i}=\frac{1}{r_{m}}\eta _{i+1}+\frac{a_{i}}{r_{m}}\varepsilon_{1},~~ i=1,2,\ldots ,n-1, \\
\dot{\eta}_{n}=-\frac{1}{r_{m}}\sum\limits_{j=1}^{n}b_{j}\eta _{j}+\frac{a_{n}}{r_{m}}\varepsilon _{1}.
\end{array}
\right.
\end{eqnarray}

Based on \eqref{xjff-3.1.1}, the compact form can be derived as follows:
\begin{eqnarray} \label{ff-1.4}
\dot{\eta}=\frac{1}{r_{m}}B\eta +\frac{1}{r_{m}}a\varepsilon _{1},~~ \text{for}  ~ t\in
\left[ t_{m},t_{m+1}\right),
\end{eqnarray}
where $\eta=\left[ \eta _{1},\eta _{2},\ldots ,\eta _{n}\right] ^{\mathrm{T}}$, $B=\Delta -\rho b^{\mathrm{T}}$, $\Delta =\bigg[
\begin{array}{cc}
\textbf{0}_{n-1} & I_{( n-1) \times ( n-1) } \\
0 & \textbf{0}_{n-1}^{\mathrm{T}}
\end{array}
\bigg] $, $\rho=\left[ \textbf{0}_{n-1}^{\mathrm{T}},1\right]^{\mathrm{T}}$, $b=[ b_{1},b_{2},\ldots ,$ $ b_{n}] ^{\mathrm{T}}$, and $a=[ a_{1},a_{2},\ldots ,a_{n}] ^{\mathrm{T}}$.

From \eqref{pf1} and \eqref{ff-1.1}-\eqref{ff-1.3}, we have that, for $t\in\left[t_{m},t_{m+1}\right) $,
\begin{eqnarray} \label{xjff-3.1.2}
\left\{
\begin{array}{l}
\dot{\varepsilon}_{i}=\frac{1}{r_{m}}\varepsilon _{i+1}-\frac{a_{i}}{r_{m}}\varepsilon _{1}+\frac{f_{i}}{
r_{m}^{n-i+1}}, ~ i=1,2,\ldots ,n-1, \\
\dot{\varepsilon}_{n}=-\frac{a_{n}}{r_{m}}\varepsilon _{1}.
\end{array}
\right.
\end{eqnarray}

Define $\varepsilon=\left[ \varepsilon _{1},\varepsilon _{2},\ldots ,\varepsilon _{n}\right] ^{\mathrm{T}}$. By \eqref{xjff-3.1.2}, the compact form can be derived as follows:
\begin{eqnarray} \label{ff-1.5}
\dot{\varepsilon}=\frac{1}{r_{m}}A\varepsilon  +\Psi,~~ \text{for}  ~ t\in
\left[ t_{m},t_{m+1}\right),
\end{eqnarray}
where $A=\Delta -a\bar{\rho} ^{\mathrm{T}}$, $\bar{\rho} =\left[ 1,\textbf{0}_{n-1}^{\mathrm{T}}\right] ^{\mathrm{T}}$, and $\Psi =\big[ \frac{f_{1}}{r_{m}^{n}},\ldots ,\frac{f_{n-1}}{r_{m}^{2}},0\big] ^{\mathrm{T}}$.

Combining \eqref{ff-1.4} and \eqref{ff-1.5} together, the closed-loop system can be expressed as follows:
\begin{eqnarray} \label{ff-1.6}
\dot{\xi}=\frac{1}{r_{m}}\Xi \xi  +\Theta,~~ \text{for}  ~ t\in \left[ t_{m},t_{m+1}\right),
\end{eqnarray}
where $\xi =\left[ \eta ^{\mathrm{T}},\varepsilon ^{\mathrm{T}}\right] ^{\mathrm{T}}$, $\Xi =\bigg[
\begin{array}{cc}
B & a\bar{\rho} ^{\mathrm{T}} \\
0_{n\times n} & A
\end{array}
\bigg] $, and $\Theta =\left[ \textbf{0}_{n}^{\mathrm{T}},\Psi ^{\mathrm{T}}\right] ^{\mathrm{T}}$.

Since $A$ and $B$ are Hurwitz matrices, one can conclude that $\Xi$ is also a Hurwitz matrix. Therefore, there exists a positive definite matrix $P\in \mathbb{R}^{2n\times 2n}$ such that
\begin{eqnarray} \label{matrix-P}
\Xi ^{\mathrm{T}}P+P\Xi \leq -I_{2n \times 2n}.
\end{eqnarray}

\subsection{Lyapunov Function Candidate} \label{Sec3.2}

Consider the following Lyapunov function candidate,
\begin{eqnarray} \label{ff-2.1}
V_{\xi }=\xi ^{\mathrm{T}}P\xi,
\end{eqnarray}
where the matrix $P$ is given in \eqref{matrix-P}.

Then, by \eqref{ff-1.6} and \eqref{ff-2.1}, we obtain
\begin{eqnarray} \label{ff-2.2}
\dot{V}_{\xi }\leq -\frac{1}{r_{m}}\Vert \xi \Vert^{2} +2\xi ^{\mathrm{T}}P\Theta.
\end{eqnarray}

Based on \eqref{pfa2}, \eqref{ff-1.2} and \eqref{ff-1.3}, one deduces that, for $i=1,2,\ldots, n-1$,
\begin{eqnarray} \label{ff-2.5}
\begin{aligned}
\frac{\vert f_{i}\vert }{r_{m}^{n-i+1}}
& \leq \frac{\theta }{r_{m}^{n-i+1}}\gamma ( u) (1+\vert y\vert ^{p}) \bigg( \sum_{j=i+2}^{n+1}\vert x_{j}\vert +\vert u\vert \bigg)  \\
& \leq \frac{c_{1}}{r_{m}^{2}}\gamma ( u) ( 1+\vert y\vert ^{p})\Vert \xi \Vert ,
\end{aligned}
\end{eqnarray}
where $x_{n+1}=0$ and $c_{1}= \theta \sqrt{n}( 2+\max\limits_{j=1,\ldots ,n}\{ b_{j}\} )$.

For the last term of \eqref{ff-2.2}, noting \eqref{ff-2.5}, we get
\begin{eqnarray} \label{ff-2.6}
\begin{aligned}
2\xi ^{\mathrm{T}}P\Theta \leq  \frac{c_{2}}{r_{m}^{2}}( \varsigma^{-1} \gamma (u) ( 1+\vert y\vert ^{p}) +\varsigma^{-1})\Vert \xi \Vert ^{2},
\end{aligned}
\end{eqnarray}
where $c_{2}=2\varsigma c_{1}( n-1)\Vert P\Vert  $ and $\varsigma$ is a known adjustable positive constant.

Then, it follows from \eqref{ff-2.2} and  \eqref{ff-2.6} that
\begin{eqnarray} \label{ff-2.8}
\dot{V}_{\xi }\leq -\frac{1}{r_{m}}\Vert \xi \Vert ^{2}+ \frac{c_{2}}{r_{m}^{2}}\vartheta (t)\Vert \xi \Vert ^{2},~~ \text{for}  ~ t\in \left[ t_{m},t_{m+1}\right) ,
\end{eqnarray}
where $\vartheta ( t) =\varsigma^{-1}\gamma ( u) ( 1+\vert y\vert ^{p}) +\varsigma^{-1}$.

Furthermore, according to \eqref{ff-1.2} and \eqref{ff-1.3}, and noting $r_{m}\geq1$, we have
\begin{eqnarray} \label{ff-2.9}
\begin{array}{l}
\gamma (u)\leq \tilde{\phi}_{1}(\vert u\vert )\leq \tilde{
\phi}_{2}(V_{\xi }), \\[1.2mm]
1+\vert y\vert ^{p}\leq 1+r_{m}^{np}\vert
\varepsilon _{1}+\eta _{1}\vert ^{p}\leq r_{m}^{np}\big(
1+c_{3}V_{\xi }^{ \frac{p}{2}}\big),
\end{array}
\end{eqnarray}
where $c_{3}= 2^{\frac{p}{2}}\lambda _{\min }^{-\frac{p}{2}}( P) $, $\tilde{\phi}_{1}(\vert u\vert )$ and $\tilde{\phi}_{2}(V_{\xi })$ are positive\,nondecreasing functions about $\vert u\vert$ and $V_{\xi }$, respectively.

Based on \eqref{ff-2.8} and \eqref{ff-2.9}, there must exist a known positive nondecreasing function $\phi ( V_{\xi })$ about $V_{\xi }$, and an unknown constant $c_{4}$, such that, for $t\in\left[t_{m},t_{m+1}\right) $,
\begin{eqnarray} \label{ff-2.10}
\dot{V}_{\xi }\leq -\frac{1}{r_{m}}\Vert \xi \Vert ^{2}+\frac{c_{4}}{r_{m}^{2}}r_{m}^{np}\phi ( V_{\xi }) \Vert \xi \Vert ^{2}.
\end{eqnarray}

\subsection{Design of Switching Mechanism} \label{utMASsgmd}

Here, the initialization and switching logic are presented.

(\uppercase\expandafter{\romannumeral1}) \emph{Initialization}:
\begin{itemize}
\item Choose three positive sequences $\bar{\sigma}=\{ \bar{\sigma}_{m},m=1,2,\ldots \} $, $\underline{\sigma }=\{ \underline{\sigma}_{m},m=1,2,\ldots \} $, and $\mu =\{ \mu_{m},m=1,2,\ldots \} $, where $\bar{\sigma}_{m}$ is monotonically increasing and satisfies $\lim\limits_{m\rightarrow +\infty }\bar{\sigma}_{m}=+\infty $, and $\underline{\sigma }_{m}$ is monotonically decreasing and satisfies $\lim\limits_{m\rightarrow +\infty }\underline{\sigma }_{m}=0$.
\item Set $r_{0}\geq 1$ and $m=1$.
\end{itemize}

(\uppercase\expandafter{\romannumeral2}) \emph{Switching Logic}:
\begin{itemize}
  \item For $t\in \left[ t_{m},t_{m+1}\right) $, the gain is set as
  \begin{equation} \label{ut3.3}
  r_{m}=\max \{ r_{m-1},\bar{\sigma}_{m}\phi ^{\frac{1}{1-np}}(\omega _{m})\} .
  \end{equation}
  \item Define
  \begin{equation} \label{ut3.1}
  \begin{array}{l}
  \chi _{m}(t)=\underline{\sigma }_{m}( \Vert \eta \Vert ^{2}+\varepsilon _{1}^{2}) +\frac{\phi ( \omega _{m}) }{r_{m}^{2-np}}\int_{t_{m}}^{t}( \Vert \eta \Vert ^{2}+\varepsilon _{1}^{2}) ds, \\[1.2mm]
  \omega _{m}=\bar{\sigma}_{m} ^{m}e^{\int_{0}^{t_{m}}\bar{\sigma}_{m}\bar{\vartheta} (s) ds},
  \end{array}
  \end{equation}
where $\bar{\vartheta}(s) = \frac{\vartheta(s)}{r_{q}^{2}}$, for $ s\in \left[ t_{q},t_{q+1}\right)$, $q=1,2,\ldots,m$.
  \item If the condition
  \begin{eqnarray} \label{ut3.4}
  \chi _{m}(t)\cdot \varphi ( \chi _{m}(t),\mu _{m}) \geq \omega_{m}
  \end{eqnarray}
  is satisfied for $t> t_{m}$, then $t$ is set as the next switching moment, that is, $t_{m+1}\leftarrow t$. Update $m\leftarrow m+1$, and repeat Step (\uppercase\expandafter{\romannumeral2}) until no such finite moment is detected.
\end{itemize}

\subsection{Main Results} \label{ut-mr}

Now, the main results are summarized in the following theorem.

\begin{theorem} \label{t1}
Consider system \eqref{pf1} under Assumption \ref{a2}. The objective of global regulation can be achieved by the SAOF controller \eqref{ff-1.3} composed of the low-gain observer \eqref{ff-1.1} and LBS gain \eqref{ut3.3}.
\end{theorem}

\begin{IEEEproof}
In view of the system dynamics subject to the inherent nonlinearity, i.e., $\vert y\vert ^{p}$, $p\in [ 0,\frac{1}{n}) $, it is usually difficult to ensure the uniqueness of the solution satisfying the local Lipschitz condition. As pointed out in \cite{LHF2022An}, global regulation of the feedforward nonlinear system describes the asymptotic behavior of its trajectory starting from an initial value. Hence, as mentioned in \cite{DSH2010Global} and \cite{QC2001A}, it suffices to analyze the existence rather than the uniqueness of the system solution. From \eqref{pfa2} and \eqref{ff-1.6}, it can be deduced that, in each switching interval $\left[ t_{m},t_{m+1}\right) $, $m=1,2,\ldots,$ the vector field of the resulting closed-loop system is continuous in its arguments. Then, similar to \cite{DSH2010Global} and \cite{QC2001A}, the resulting closed-loop system has at least one solution on $\left[t_{m},t_{m,f}\right) $ with $t_{m,f}\leq t_{m+1}$. Based on \eqref{ut3.1} and \eqref{ut3.4}, it follows that $\|\eta \|$ and $\varepsilon _{1}$ are bounded on $\left[ t_{m},t_{m+1}\right) $. Clearly, by \eqref{ff-1.2} and \eqref{ff-1.3}, we have $\frac{\vartheta (t)}{r^2_{m}}$ is also bounded on $\left[ t_{m},t_{m+1}\right)$. For this, \eqref{ff-2.8} is further expressed as $\dot{V}_{\xi }\leq c^{\ast}V_{\xi }$, $t\in \left[ t_{m},t_{m+1}\right) $, where $c^{\ast }$ is a positive constant. Hence, the system solution exists in each switching interval. In what follows, let $\left[ 0,t_{f}\right) $ be the interval for the solution of the closed-loop system.

By \eqref{ff-2.8}, we deduce
\begin{eqnarray} \label{ut4.1}
\dot{V}_{\xi }(t)\leq \frac{c _{2}\vartheta (t)}{r_{m}^{2}\lambda _{\min }(P) }V_{\xi }(t),~~ \text{for}  ~ t\in \left[ t_{m},t_{m+1}\right) .
\end{eqnarray}

Integrating \eqref{ut4.1} from $t_{m}$ to $t_{m+1}^{-}$, one has
\begin{eqnarray} \label{ut4.2}
V_{\xi }( t_{m+1}^{-}) \leq e^{\int_{t_{m}}^{t_{m+1}}c_{5} \bar{\vartheta} (s)ds}V_{\xi }( t_{m}) ,
\end{eqnarray}
where $c_{5}=\frac{c_{2}}{\lambda _{\min }( P) }$.

Since $r_{m}\leq r_{m+1}$ and $x_{i}$, $\hat{x}_{i}$ are continuous, one can obtain that $\frac{V_{\xi }( t_{m+1}) }{\lambda _{\max }( P) }\leq \Vert \xi ( t_{m+1}) \Vert ^{2}\leq \Vert \xi ( t_{m+1}^{-}) \Vert ^{2}\leq \frac{V_{\xi }( t_{m+1}^{-}) }{\lambda _{\min }( P) }$. It then follows that $V_{\xi }(t_{m+1})\leq \beta V_{\xi }(t_{m+1}^{-})$, where  $\beta =\frac{\lambda _{\max }( P) }{\lambda _{\min }(P) }$. Then, by \eqref{ut4.2}, it is straightforward to see that
\begin{eqnarray} \label{ut4.3}
V_{\xi }( t_{m+1}) \leq \beta e^{\int_{t_{m}}^{t_{m+1}}c_{5} \bar{\vartheta} (s)ds}V_{\xi }( t_{m}) .
\end{eqnarray}

After iterative calculation, it follows from \eqref{ut4.3} that
\begin{eqnarray} \label{ut4.4}
\begin{aligned}
V_{\xi }(t_{m}) & \leq \beta ^{m-1}e^{\int_{0}^{t_{m}} c _{5} \bar{\vartheta} ( s) ds} V_{\xi }( 0)\\
& \leq \alpha ^{m}e^{\int_{0}^{t_{m}}\alpha \bar{\vartheta} ( s) ds},
\end{aligned}
\end{eqnarray}
where $\alpha =\max \{\beta, c _{5},V_{\xi }( 0)\} $.

Next, suppose that there exist infinite switching moments. Hence, there exists a large positive integer $m^{\ast }$ satisfying
\begin{eqnarray} \label{ut4.5}
\bar{\sigma}_{m^{\ast }}^{1-np} > c _{4}+ 1, ~ \omega _{m^{\ast }} >
V_{\xi }( t_{m^{\ast }}) , ~ \underline{\sigma }
_{m^{\ast }} < \lambda _{\min }( P) .
\end{eqnarray}

We then analyze the properties of the function on $[ t_{m^{\ast }},t_{\bar{m}}) $, where $t_{\bar{m}}\in ( t_{m^{\ast }},t_{m^{\ast }+1}] $ denotes a maximum moment satisfying $\omega _{m^{\ast }}\geq V_{\xi }(t) $. By \eqref{ff-2.10}, \eqref{ut3.3} and \eqref{ut4.5}, one obtains that, for $t\in [ t_{m^{\ast }},t_{\bar{m}}) $,
\begin{eqnarray} \label{ut4.6}
\begin{aligned}
\dot{V}_{\xi } (t) & \leq -\frac{1}{r_{m^{\ast }}^{2-np}}\left( r_{m^{\ast }}^{1-np}-c_{4}\phi
( V_{\xi }( t) ) \right) \Vert \xi \Vert ^{2} \\
& \leq -\frac{1}{r_{m^{\ast }}^{2-np}}\left( \bar{\sigma}_{m^{\ast }}^{1-np}\phi
( \omega _{m^{\ast }}) -c_{4}\phi ( V_{\xi }(
t) ) \right) \Vert \xi \Vert ^{2} \\
& \leq -\frac{1}{r_{m^{\ast }}^{2-np}}\big( \bar{\sigma}_{m^{\ast
}}^{1-np}-c_{4}\big) \phi ( \omega _{m^{\ast }}) \Vert \xi \Vert ^{2} \\
& < -\frac{1}{r_{m^{\ast }}^{2-np}}\phi ( \omega _{m^{\ast }}) ( \Vert \eta \Vert ^{2}+\varepsilon_{1}^{2}).
\end{aligned}
\end{eqnarray}

Then, noting that $V_{\xi }(t)$ is nonincreasing on $ \left[t_{m^{\ast }},t_{\bar{m}}\right) $, one gets $t_{\bar{m}}=t_{m^{\ast}+1}$. Based on \eqref{ut4.6}, one has that, for $t\in \left[ t_{m^{\ast }},t_{m^{\ast}+1}\right) $,
\begin{eqnarray} \label{ut4.7}
V_{\xi }(t) -V_{\xi }( t_{m^{\ast }})
< -\frac{\phi ( \omega _{m^{\ast }})}{r_{m^{\ast }}^{2-np}}
\int_{t_{m^{\ast }}}^{t}( \Vert \eta \Vert ^{2}+\varepsilon
_{1}^{2}) ds.
\end{eqnarray}

By \eqref{ut4.5} and \eqref{ut4.7}, it can be inferred that
\begin{eqnarray} \label{ut4.8}
\begin{aligned}
& \underline{\sigma }_{m^{\ast }}( \Vert \eta \Vert
^{2}+\varepsilon_{1}^{2}) -\omega
_{m^{\ast }}  < \lambda _{\min }\left( P\right) \left\Vert \xi \right\Vert ^{2}-\omega
_{m^{\ast }} \\
& < -\frac{\phi ( \omega _{m^{\ast }})}{r_{m^{\ast }}^{2-np}}
\int_{t_{m^{\ast }}}^{t}( \Vert \eta \Vert ^{2}+\varepsilon_{1}^{2}) ds.
\end{aligned}
\end{eqnarray}

From \eqref{ut3.1} and \eqref{ut4.8}, we obtain
\begin{eqnarray} \label{ut4.9}
\chi _{m^{\ast }}(t)  < \omega _{m^{\ast }}.
\end{eqnarray}

Hence, it follows from \eqref{ut4.9} that
\begin{eqnarray} \label{jut4.9}
\chi _{m^{\ast }}(t)\cdot \varphi ( \chi _{m^{\ast }}(t),\mu_{m^{\ast }}) < \omega _{m^{\ast }}.
\end{eqnarray}

By \eqref{jut4.9}, we know that condition \eqref{ut3.4} is no longer satisfied as well as the hypothesis does not hold. Thus, we arrive at the finiteness of the switching moment.

Next, we denote the last switching moment as $t_{M}$. Then, after moment $t_{M}$, the following inequality holds,
\begin{eqnarray} \label{ut4.2.1}
\chi _{M}(t)\cdot \varphi ( \chi _{M}(t),\mu _{M}) < \omega _{M}.
\end{eqnarray}

Applying Lemma \ref{srf-lemma} to \eqref{ut4.2.1}, one deduces
\begin{eqnarray} \label{ut4.2.2}
\chi _{M}(t)-0.2785\mu _{M} < \omega _{M}.
\end{eqnarray}

It then follows from \eqref{ut3.1} and \eqref{ut4.2.2} that
\begin{eqnarray} \label{ut4.2.3}
\begin{aligned}
& \underline{\sigma }_{M}( \Vert \eta \Vert ^{2}+\varepsilon _{1}^{2}) +\frac{\phi ( \omega _{M}) }{r_{M}^{2-np}}\int_{t_{M}}^{t}( \Vert \eta \Vert ^{2}+\varepsilon _{1}^{2}) ds \\
& < \omega _{M}+0.2785\mu_{M}.
\end{aligned}
\end{eqnarray}

From \eqref{ut4.2.3}, and the boundedness of $\underline{\sigma }_{M}$, $\omega _{M}$, $\phi (\omega _{M})$, $r _{M}$ and $\mu _{M}$, we obtain that $\Vert \eta \Vert $, $\varepsilon_{1}$, $\int_{t_{M}}^{t}\Vert \eta \Vert^{2}ds$ and $\int_{t_{M}}^{t}\varepsilon _{1}^{2}ds$ are bounded on $\left[ t_{M},t_{f}\right) $, where $t_{f}$ is the maximum upper bound of the solution interval. Therefore, it further yields that $u$, $y$, $\hat{x}_{i}$ and $\int_{t_{M}}^{t}\hat{x}_{i}^{2}ds$ are bounded on $\left[ t_{M},t_{f}\right) $.

Next, we analyze the boundedness of $\Vert \varepsilon\Vert $ and $ \int_{t_{M}}^{t}\Vert \varepsilon\Vert ^{2}ds$. For $t\in \left[0,t_{f}\right) $, the following state transformation is introduced,
\begin{eqnarray} \label{ut4.2.4}
\zeta _{i}=\frac{x_{i}-\hat{x}_{i}}{\bar{r}_{M}^{n-i+1}},~~ i=1,2,\ldots ,n,
\end{eqnarray}
where $\bar{r}_{M}=\max \left\{ r_{M},\kappa_{4}+1\right\} $ with $\kappa_{4}$ being an unknown positive constant to be determined later.

Then, we focus on $t\in \left[t_{M},t_{f}\right) $. Combining \eqref{pf1}, \eqref{ff-1.1}-\eqref{ff-1.3} and \eqref{ut4.2.4} together, one has
\begin{eqnarray} \label{ut4.2.5}
\dot{\zeta}=\frac{1}{\bar{r}_{M}} A\zeta +\frac{1}{\bar{r}_{M}}a\zeta_{1}-\frac{1}{\bar{r}_{M}}Ra\varepsilon_{1}  +\bar{\Psi},
\end{eqnarray}
where $\zeta=\left[ \zeta _{1},\zeta _{2},\ldots ,\zeta _{n} \right] ^{\mathrm{T}}$, $R=\textrm{diag}\big\{ \frac{r_{M}^{n-1}}{\bar{r}_{M}^{n-1}},\ldots ,\frac{r_{M}}{\bar{r}_{M}},1\big\} $, and $\bar{\Psi}=\big[ \frac{f_{1}}{\bar{r}_{M}^{n}},\ldots ,\frac{f_{n-1}}{\bar{r}_{M}^{2}},0\big] ^{\mathrm{T}}$.

For system \eqref{ut4.2.5}, a Lyapunov function candidate is selected as
\begin{eqnarray} \label{ut.A.1}
V_{\zeta }=\zeta ^{\mathrm{T}}Q\zeta , ~~ \text{for}  ~ t\in \left[t_{M},t_{f}\right) ,
\end{eqnarray}
where $Q \in\mathbb{R}^{n \times n }$ is a positive definite matrix satisfying $A ^{\mathrm{T}}Q+QA \leq -I_{n}$.

Then, by \eqref{ut4.2.5} and \eqref{ut.A.1}, we have that, for $t\in \left[t_{M},t_{f}\right) $,
\begin{eqnarray} \label{ut.A.2}
\begin{aligned}
\dot{V}_{\zeta }\leq & -\frac{1}{\bar{r}_{M}}\Vert \zeta \Vert^{2}+\frac{2}{\bar{r}_{M}}\zeta ^{\mathrm{T}}Qa\zeta _{1}-\frac{2}{\bar{r}_{M}}\zeta^{\mathrm{T}}QRa\varepsilon _{1} \\
& +2\zeta ^{\mathrm{T}}Q\bar{\Psi}.
\end{aligned}
\end{eqnarray}

Based on Young's inequality, we get
\begin{eqnarray} \label{ut.A.3}
\begin{array}{l}
\frac{2}{\bar{r}_{M}}\zeta ^{\mathrm{T}}Qa\zeta _{1}\leq \frac{1}{\bar{r}_{M}^{2}}
\Vert \zeta \Vert ^{2}+\kappa_{1}\zeta _{1}^{2}, \\[1mm]
-\frac{2}{\bar{r}_{M}}\zeta ^{\mathrm{T}}QRa\varepsilon _{1}\leq \frac{1}{\bar{r}
_{M}^{2}}\Vert \zeta \Vert ^{2}+\kappa_{1}\varepsilon _{1}^{2},
\end{array}
\end{eqnarray}
where $\kappa _{1}=\max \{\Vert Qa\Vert ^{2},\Vert QRa\Vert^{2}\}$.

Moreover, \eqref{ut4.2.3} implies that $u$ and $y$ are bounded on $\left[ t_{M},t_{f}\right) $. Then, it follows from \eqref{pfa2}, \eqref{ff-1.2} and \eqref{ut4.2.4} that
\begin{eqnarray} \label{ut.A.5}
\begin{aligned}
\frac{\vert f_{i}\vert }{\bar{r}_{M}^{n-i+1}}
& \leq \frac{\theta }{\bar{r}_{M}^{n-i+1}}\gamma(u)(1+\vert y\vert ^{p}) \bigg( \sum_{j=i+2}^{n+1}\vert x_{j}\vert +\vert u\vert \bigg)  \\
& \leq \frac{\kappa_{2}}{\bar{r}_{M}^{2}}\left( \Vert \zeta \Vert +\Vert \eta \Vert  \right),~~ i=1,2,\ldots, n-1,
\end{aligned}
\end{eqnarray}
where $\kappa _{2}=\theta \sqrt{n}\big( 1+\max\limits_{i=1,\ldots,n}\{b_{i}\}\big) \sup\limits_{t\in [ t_{M},t_{f}) }\{\gamma (u) (1+\vert y\vert ^{p})\}$.

According to \eqref{ut.A.5}, we obtain
\begin{eqnarray} \label{ut.A.6}
\begin{aligned}
2\zeta ^{\mathrm{T}}Q\bar{\Psi}\leq \frac{\kappa_{3}}{\bar{r}_{M}^{2}}\left( \Vert \zeta \Vert ^{2}+\Vert \eta \Vert ^{2}\right),
\end{aligned}
\end{eqnarray}
where $\kappa_{3}=3(n-1)\Vert Q\Vert \kappa _{2}$.

Next, by \eqref{ut.A.2}, \eqref{ut.A.3} and \eqref{ut.A.6} and noting $\bar{r}_{M}\geq \kappa_{4}+1$, we have that, for $t\in \left[t_{M},t_{f}\right) $,
\begin{eqnarray} \label{ut.A.8}
\begin{aligned}
\dot{V}_{\zeta }& \leq -\frac{\bar{r}_{M}-\kappa_{4}}{\bar{r}_{M}^{2}}\Vert \zeta
\Vert ^{2}+\kappa_{4}( \zeta _{1}^{2}+\varepsilon _{1}^{2}+\Vert \eta \Vert ^{2}) \\
& \leq -\frac{1}{\bar{r}_{M}^{2}}\Vert \zeta \Vert
^{2}+\kappa_{4}( \zeta _{1}^{2}+\varepsilon _{1}^{2}+\Vert \eta \Vert ^{2}),
\end{aligned}
\end{eqnarray}
where $\kappa _{4}=2+\kappa _{1}+\kappa _{3}$.

Integrating both sides of \eqref{ut.A.8}, we arrive at, for $t\in \left[t_{M},t_{f}\right) $,
\begin{eqnarray} \label{ut.A.9}
\begin{aligned}
& \lambda _{\min }( Q) \Vert \zeta \Vert ^{2}+\frac{1}{\bar{r}_{M}^{2}}\int_{t_{M}}^{t}\Vert \zeta \Vert ^{2}ds \\
& ~\leq \kappa _{4}\int_{t_{M}}^{t}\big( \zeta _{1}^{2}+\varepsilon _{1}^{2}+\Vert \eta \Vert ^{2}\big)ds +V_{\zeta }(t_{M}).
\end{aligned}
\end{eqnarray}

Since $\int_{t_{M}}^{t}\varepsilon_{1}^{2}ds$ and $ \int_{t_{M}}^{t}\Vert \eta \Vert ^{2}ds$ are bounded on $\left[ t_{M},t_{f}\right) $, we can deduce that $\int_{t_{M}}^{t}\big( \zeta _{1}^{2}+\varepsilon_{1}^{2}+\Vert \eta \Vert ^{2}\big)ds$ is bounded on $\left[ t_{M},t_{f}\right) $. From \eqref{ut.A.9}, we obtain that $\Vert \zeta \Vert $ and $\int_{t_{M}}^{t}\Vert \zeta \Vert ^{2}ds$ are bounded on $\left[ t_{M},t_{f}\right) $.

According to \eqref{ff-1.2} and \eqref{ut4.2.4}, it is easy to verify that $\Vert\varepsilon\Vert $ and $\int_{t_{M}}^{t}\Vert \varepsilon\Vert ^{2}ds$ are bounded on $\left[ t_{M},t_{f}\right) $. Based on the above analysis and the continuity of $x_{i}$ and $\hat{x}_{i}$, we obtain that all signals of system \eqref{ff-1.6} are bounded on $\left[ 0,t_{f}\right) $. Hence, $t_{f}$\ can be extended to its maximum, i.e., $t_{f}=+\infty $. From \eqref{ff-1.4} and \eqref{ff-1.5}, we have that $\Vert\dot{\eta}\Vert $ and $\Vert\dot{\varepsilon}\Vert $ are bounded on $\left[ 0,+\infty \right) $. Based on Barbalat lemma \cite{SJJ1991Applied}, we can deduce that $\lim\limits_{t\rightarrow +\infty }\Vert\eta\Vert=0$ and$\lim\limits_{t\rightarrow +\infty }\Vert\varepsilon\Vert=0$. Therefore, we can conclude that $\lim\limits_{t\rightarrow +\infty } x_{i}=0$ and $\lim\limits_{t\rightarrow +\infty } \hat{x}_{i}=0$, $i=1,2,\ldots ,n$.
\end{IEEEproof}

\begin{remark} \rm \label{XJ-02}
In this article, the positive sequences $\bar{\sigma}$, $\underline{\sigma }$ and $\mu$ play a crucial role in improving the control performance. Generally, the positive sequence $\underline{\sigma }$ should be chosen as a set of slowly decreasing positive constants to enhance the sensitivity to the unknown constants. The positive sequence $\bar{\sigma}$ should be chosen as a set of slowly increasing positive constants to avoid overestimating the unknown constants. The positive sequence $\mu$ should be chosen as a set of suitably large positive constants to improve convergence speed and transient performance. In addition, the selection of the function $\phi (\cdot)$ is not unique. A larger function $\phi (\cdot)$ may result in larger candidate switching gains, which in turn may lead to slow convergence speed and poor transient performance.
\end{remark}

\section{Discussion on Disturbance Tolerance} \label{Sec4}

This section is devoted to discussing whether the proposed LBS gain approach can be applied to system \eqref{pf1} with unknown external disturbances. In this case, consider the following system:
\begin{eqnarray} \label{s4.1}
\left\{
\begin{array}{l}
\dot{x}_{i}=x_{i+1}+f_{i}(t,x,u)+\varpi _{i}(t), \\
\dot{x}_{n}=u+\varpi _{n}(t), ~~ i=1,2,\ldots ,n-1, \\
y=x_{1},
\end{array}
\right.
\end{eqnarray}
where $\varpi _{i}( t) ,$ $i=1,2,\ldots ,n-1,$ and $\varpi_{n}( t) $ are unknown external disturbances and the remaining system variables are the same as those shown in system \eqref{pf1} while satisfying Assumption \ref{a2}. Similar to \cite{HY2019Switching,MG2019Otli,FDB2023Grol}, we make the following assumption for the external additive disturbances.

\begin{assumption} \rm \label{a3}
For $i=1,2,\ldots ,n,$ there exists an unknown nonnegative constant $\varpi ^{\ast }$, such that
\begin{eqnarray*}
\sup\limits_{t\in [ 0,+\infty ) }\bigg( \vert \varpi _{i}( t) \vert +\int_{0}^{t}\varpi_{i}^{2}(s) ds\bigg) \leq \varpi ^{\ast }.
\end{eqnarray*}
\end{assumption}

For system \eqref{s4.1} under Assumptions \ref{a2} and \ref{a3}, an improved control approach is designed as follows.

Similar to Subsection \ref{Sec3.1}, after the same state transformation \eqref{ff-1.2}, one has, for $t\in\left[t_{m},t_{m+1}\right) $,
\begin{eqnarray*}
\dot{\xi}=\frac{1}{r_{m}}\Xi \xi +\Theta +\frac{1}{r_{m}}\Gamma ,
\end{eqnarray*}
where $\Xi $ and $\Theta $ are the same as those in system \eqref{ff-1.6}, $\Gamma =\big[ \textbf{0}_{n}^{\mathrm{T}},\frac{\varpi _{1}}{r_{m}^{n-1}},\ldots , \frac{\varpi _{n-1}}{r_{m}},\varpi _{n}\big] ^{\mathrm{T}}$.

By selecting the Lyapunov function candidate in the same way as in Subsection \ref{Sec3.2}, one deduces that, for $t\in \left[ t_{m},t_{m+1}\right) $,
\begin{eqnarray*}
\dot{V}_{\xi }\leq -\frac{1}{2r_{m}}\Vert \xi \Vert ^{2}+ \frac{c_{2}}{r_{m}^{2}}\vartheta ( t) \Vert \xi \Vert^{2}+\frac{2}{r_{m}}\Vert P\Vert ^{2}\Vert \Gamma \Vert ^{2},
\end{eqnarray*}
where $c_{2}$, $ P $, and $\vartheta ( t) $ are the same as those in \eqref{ff-2.8}.

It then follows that, for $t\in \left[ t_{m},t_{m+1}\right) $,
\begin{eqnarray*}
\dot{V}_{\xi }\leq -\frac{1}{2r_{m}}\Vert \xi \Vert ^{2}+ \frac{c_{4}}{r_{m}^{2}}r_{m}^{np}\phi ( V_{\xi }) \Vert \xi \Vert ^{2}+\frac{2}{r_{m}}\Vert P\Vert ^{2}\Vert \Gamma \Vert ^{2},
\end{eqnarray*}
where $c_{4}$ and $\phi ( V_{\xi }) $ are selected in the same way as in \eqref{ff-2.10}.

After the same initialization as in Subsection \ref{utMASsgmd}, we improve the switching logic as follows:
\begin{itemize}
  \item For $t\in \left[ t_{m},t_{m+1}\right) $, the gain is set as
  \begin{equation} \label{szy4.2}
  r_{m}=\max \{ r_{m-1},2^{\frac{1}{1-np}}\bar{\sigma}_{m}\phi ^{\frac{1}{1-np}}(\omega _{m})\} .
  \end{equation}
  \item Define
  \begin{equation} \label{szy4.3}
  \begin{array}{l}
  \chi _{m}(t)=\underline{\sigma }_{m}( \Vert \eta \Vert ^{2}+\varepsilon _{1}^{2}) +\frac{\phi ( \omega _{m}) }{r_{m}^{2-np}}\int_{t_{m}}^{t}( \Vert \eta \Vert ^{2}+\varepsilon _{1}^{2}) ds, \\[1.2mm]
  \omega _{m}=\bar{\sigma}_{m} ^{m}e^{\int_{0}^{t_{m}}\bar{\sigma}_{m}\bar{\vartheta} ( s) ds}( t_{m}+1),
  \end{array}
  \end{equation}
where $\bar{\vartheta}(s)$ is in the same form as defined in \eqref{ut3.1}.
  \item If the condition
  \begin{eqnarray} \label{szy4.4}
  \chi _{m}(t)\cdot \varphi ( \chi _{m}(t),\mu _{m}) \geq \omega_{m}+1
  \end{eqnarray}
  is satisfied for $t> t_{m}$, then $t$ is set as the next switching moment, that is, $t_{m+1}\leftarrow t$. Update $m\leftarrow m+1$, and repeat the above process until no such finite moment is detected.
\end{itemize}

Now, the main results under the improved control approach are summarized below.

\begin{theorem} \label{t2}
Consider system \eqref{s4.1} under Assumptions \ref{a2} and \ref{a3}. The objective of global regulation can be achieved by the SAOF controller \eqref{ff-1.3} composed of the low-gain observer \eqref{ff-1.1} and improved LBS gain \eqref{szy4.2}.
\end{theorem}

\begin{IEEEproof}
The overall proof is similar to that of Theorem \ref{t1}, and thus omitted.
\end{IEEEproof}

\begin{remark} \rm \label{XJ-02}
In this article, we use a low-gain linear observer to estimate the unmeasurable system states. From the finiteness of the switching moments, we obtain that the switching gain and states $\eta_{i}$ are bounded. Then, by state transformation \eqref{ut4.2.4} and Barbalat lemma, we derive the convergence of the observer errors. The proposed approach involves neither a tedious iterative design process nor complex estimators or observers. Therefore, the proposed approach has certain advantages in terms of simplicity and flexibility.
\end{remark}

\begin{remark} \rm \label{u-controller}
Under the control framework of this article, the advantages of the controller are mainly reflected in the following three aspects. (i) \emph{Low complexity}. Compared with \cite{LHF2022An,ZXF2011Design,ZXF2011Feedback,WP2019Global}, the controller has low complexity since it does not use the derivative of gain. (ii) \emph{Concise form}. The controller has a linear-like form, which is more concise than that in \cite{YX1999Global}. (iii) \emph{Strong adaptability}. Different from the time-varying gain approach in \cite{LHF2021Leader-follower,LQR2020Global}, the logic unit detects and supervises the dynamic behaviors in real-time, thus endowing the designed controller with stronger adaptive capability.
\end{remark}

\begin{remark} \rm \label{u-approach}
The unified framework established by the proposed approach can provide a new perspective for control problems of other nonlinear systems, for example, the global output feedback control problem of strict-feedback systems \cite{CCC2018Global} and the decentralized control problem of large-scale systems \cite{ZX2015Adaptive}. Notably, the proposed approach can also improve the control performance of strict-feedback nonlinear systems. In such systems, the peaking phenomenon caused by relatively high gain often occurs in control input, see \cite{LK2020Output}. From a practical viewpoint, this phenomenon is undesirable. We believe that our approach can achieve satisfactory control performance with a relatively small control effort.
\end{remark}

\section{Examples and Comparisons} \label{Examples}

To demonstrate the effectiveness and advantages of the LBS gain approach, representative examples and detailed comparisons are given below.

\begin{figure}[tbp]
\par
\begin{center}
\includegraphics[height=4.0cm, width=3.4in]{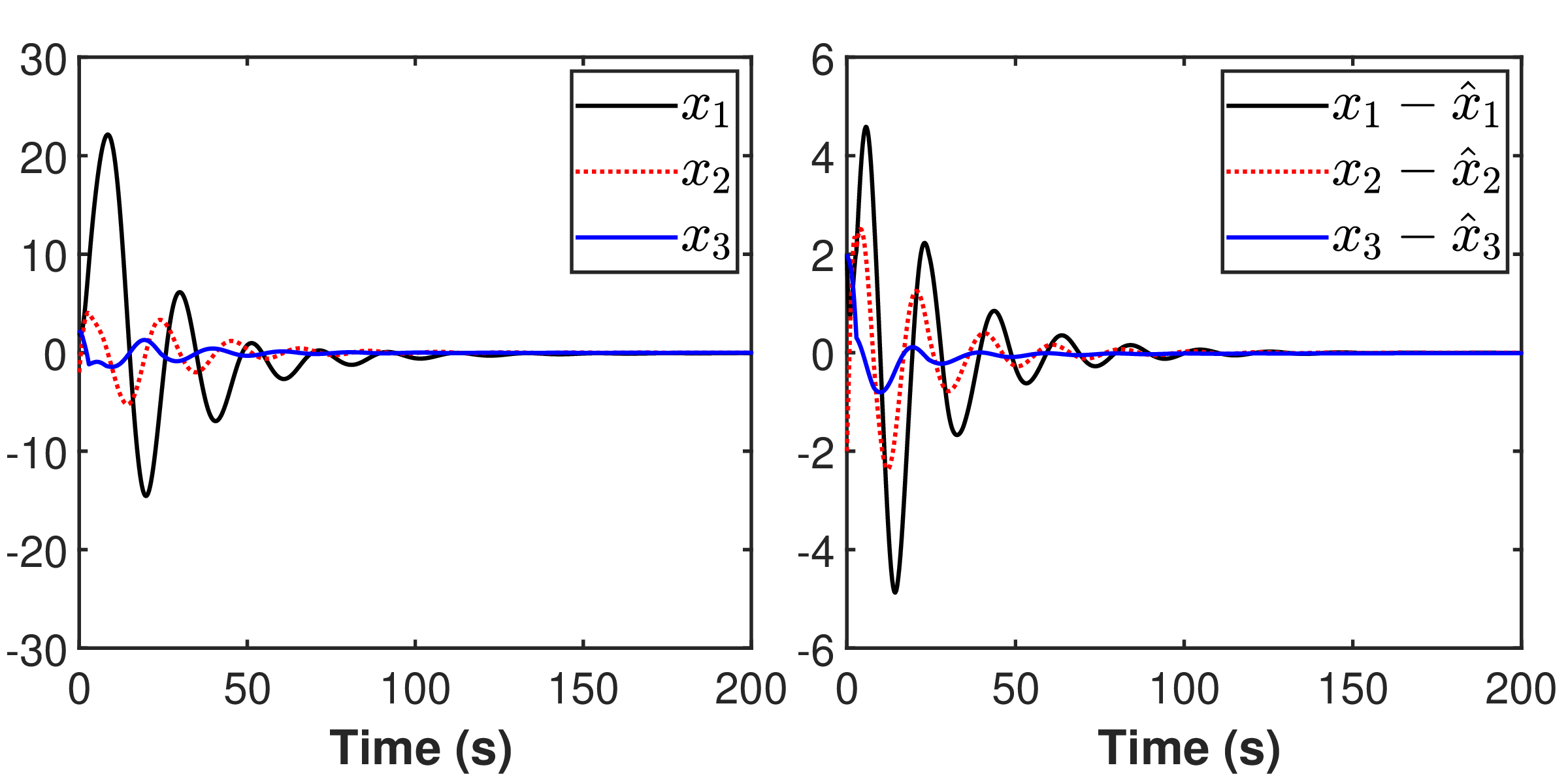}
\end{center}
\par
\caption{The trajectories of $x_{i}$ and $x_{i}-\hat{x}_{i}$, $i=1,2,3$, in Example \ref{example1}.}
\label{Fig-E1-1}

\par
\begin{center}
\includegraphics[height=4.0cm, width=3.4in]{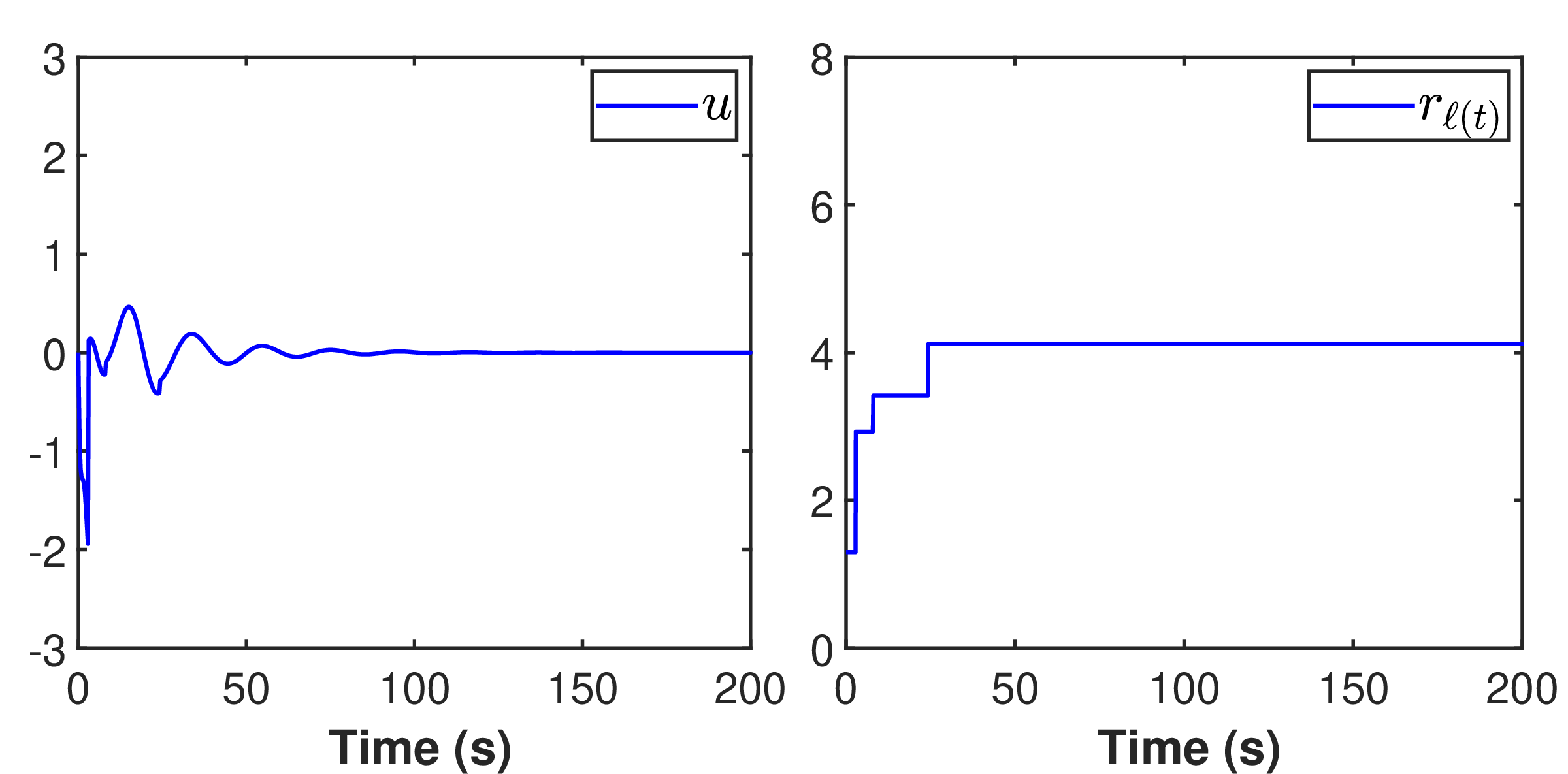}
\end{center}
\par
\caption{The trajectories of $u$ and $r_{\ell(t)}$ in Example \ref{example1}.}
\label{Fig-E1-2}
\end{figure}

\begin{example} \label{example1} \rm
Consider the following feedforward nonlinear system:
\begin{eqnarray} \label{E1-1}
\left\{
\begin{array}{l}
\dot{x}_{i}= x_{i+1}+f_{i}(\cdot )+\varpi _{i}(t), ~~ i=1,2,\\
\dot{x}_{3}= u+\varpi _{3}(t), \\
y=x_{1},
\end{array}
\right.
\end{eqnarray}
where $f_{1}(\cdot )=\theta e^{u}\ln \big( 2+\vert y\vert ^{\frac{1}{4}}\big) \Big(\frac{x_{3}\sin (x_{2})}{( 1-0.1x_{1}) ^{2}+e^{x_{3}^{2}}}+\frac{u}{1+\arctan x_{1}^{2}}\Big) $, $f_{2}(\cdot )=\theta e^{u}\ln \big( 2+\vert y\vert ^{\frac{1}{4}}\big) u$, $\varpi _{1}(t)=e^{-t},$ $\varpi _{2}(t)=\frac{2}{\sqrt{1+t^2}}$, $\varpi _{3}(t)=e^{-2t}$, and $\theta$ is an unknown nonnegative constant. For a given nonlinear system, we need to determine whether the nonlinear functions satisfy the form of \eqref{pfa2} in Assumption \ref{a2}. Obviously, we can verify that the system \eqref{E1-1} satisfies Assumption \ref{a2} with $\gamma ( u) =e^{u}$ and $p=\frac{1}{4}$.

In this simulation, the unknown constant is selected as $\theta =0.2$. The control parameters are set as $\left[a_{1},a_{2},a_{3}\right] ^{\mathrm{T}}=\left[ 1.2,1.5,1.3\right] ^{\mathrm{T}}$ and $\left[b_{1},b_{2},b_{3}\right] ^{\mathrm{T}}=\left[ 0.4,1.8,1.2\right] ^{\mathrm{T}}$. Then, we have $\lambda _{\min }( P) =0.3197$, $\lambda _{\max}( P) =96.3207$, and $c_{3}=1.2576$. Particularly, we choose $\phi ( \omega _{m}) =\varsigma^{-1}\gamma ( ( \max\limits_{i=1,2,3}\{ b_{i}\} ) \sqrt{3} \lambda _{\min }^{-0.5}( P) \omega _{m}^{\frac{1}{2}})( 1+c_{3}\omega _{m}^{\frac{p}{2}}) +\varsigma^{-1}=2.8e^{5.5139\omega _{m}^{0.5}}( 1+1.2576\omega _{m}^{0.125}) +2.8$. Moreover, the positive sequence $\bar{\sigma}$ is selected as $\bar{\sigma}_{m}=6m\times10^{-5}$, $m=1,\ldots ,$ the positive sequence $\underline{\sigma }$ is selected as $\underline{\sigma }_{m}=e^{-m}$, $m=1,\ldots ,$ the positive sequence $\mu $ is selected as $\mu _{1}=e^{8}$ $\mu _{m}=e^{3}$, $ m=2,\ldots$. The initial values are chosen as $r_{0}=1.3$, $\left[ x_{1}(0),x_{2}(0),x_{3}(0)\right] ^{\mathrm{T}}=\left[ 2,-2,2\right] ^{\mathrm{T}}$ and $\left[ \hat{x}_{1}(0),\hat{x}_{2}(0),\hat{x}_{3}(0)\right] ^{\mathrm{T}}=\left[ 0,0,0\right] ^{\mathrm{T}}$.

Under the proposed controller, the simulation results are shown in Figs. \ref{Fig-E1-1} and \ref{Fig-E1-2}. It can be seen from Figs. \ref{Fig-E1-1} and \ref{Fig-E1-2} that global regulation is achieved and trajectories of $u$ and $r_{\ell(t)}$ are bounded. To sum up, the simulation results demonstrate the effectiveness of the proposed approach.

Notably, the existing gain scaling approaches (see, e.g., \cite{LHF2022An,ZXF2011Design,LHF2021Leader-follower,ZCH2014Leader-follower,LQR2020Global}) cannot be applied to such a system \eqref{E1-1}. To make a more intuitive comparison with the existing approaches, we further consider a universal application example to verify the advantages of the proposed approach.
\end{example}

\begin{example} \label{example2} \rm
Consider a nonlinear liquid level control resonant circuit system \cite{LQR2020Global}:
\begin{eqnarray} \label{E2-1}
\left\{
\begin{array}{l}
\dot{i}_{L_{1}}=-\frac{1}{L_{1}}V_{C}-\frac{1}{L_{1}}R_{1}\left( i_{L_{2}}-\frac{1}{2}\sin V_{C}\right) , \\
\dot{V}_{C}=\frac{1}{C}i_{L_{2}}-\frac{1}{2C}\sin V_{C}, \\
\dot{i}_{L_{2}}=-\frac{1}{L_{2}}R_{2}i_{L_{2}}+\frac{1}{L_{2}}\upsilon ,
\end{array}
\right.
\end{eqnarray}
where $L_{1}$ and $L_{2}$ represent the inductances; $i_{L_{1}}$ and $i_{L_{2}}$ are the currents through $L_{1}$ and $L_{2}$; $R_{1}$ and $R_{2}$ are the resistances around $L_{1}$ and $L_{2}$, respectively; $\upsilon $ is the control voltage; $V_{C}$ represents the voltage across the capacitor $C$.

Similar to \cite{LQR2020Global}, we choose $L_{1}=L_{2}=R_{2}=1$, $C=2$ and define $R_{1}=\theta_R $, $x_{1}=i_{L_{1}}$, $x_{2}=-V_{c}$, $x_{3}=-\frac{1}{2}\left( i_{L_{2}}-\frac{1}{2}\sin V_{c}\right) $, and $u=\frac{1}{2}i_{L_{2}}-\frac{1}{2}\upsilon +\left( \frac{1}{8}i_{L_{2}}-\frac{1}{16}\sin V_{c}\right) \cos V_{c}$. Then, the system \eqref{E2-1} can be transformed into a system satisfying Assumption \ref{a2}, i.e.,
\begin{eqnarray} \label{E2-2}
\left\{
\begin{array}{l}
\dot{x}_{1}=x_{2}+2\theta_R x_{3}, \\
\dot{x}_{2}=x_{3}, \\
\dot{x}_{3}=u.
\end{array}
\right.
\end{eqnarray}

It follows from \eqref{E2-2} that Assumption \ref{a2} has reasonable practicality. To illustrate the advantages of the LBS gain approach, several cases are considered in Table \ref{C-c-a}. When $\theta_R$ is a known constant, the control approach in Case 1 or Case 2 is sufficient to counteract the minor system uncertainty. Otherwise, it is necessary to adopt other control approaches.

\begin{table}[t]
	\caption{Comparisons of different control approaches with known or unknown constant $\theta_R$.}
    \label{C-c-a}
	\centering
    \renewcommand\arraystretch{1.1}
	\setlength{\tabcolsep}{1.50mm}{
	\begin{tabular}{lll}
		\toprule[1pt]
		Case & Reference & \tabincell{l}{Control approach} \\
		\midrule[0.5pt]
		1 & \cite{LHF2021Leader-follower,ZCH2014Leader-follower}& \tabincell{l}{Static gain: \\ $r=80$} \\
		\midrule[0.5pt]
		2 &\cite{ZXF2011Design,LHF2021Leader-follower} & \tabincell{l}{Time-varying gain (bounded type): \\ $\dot{r}=\frac{1}{1.4r}\max \left\{ 80-r,0\right\} $  \\ with $r(0)=1$}   \\
		\midrule[0.5pt]
		3 & \cite{LQR2020Global} & \tabincell{l}{Time-varying gain (unbounded type):\\ $r(t)=\ln \left( t+6\right) $}  \\
		\midrule[0.5pt]
		4 & \cite{LHF2022An}	& \tabincell{l}{Dynamic gain: \\ $\dot{r}=\frac{\left( x_{1}-\hat{x}_{1}\right) ^{2}}{r^{7}}+\sum\limits_{i=1}^{3}\frac{\hat{x}_{i}^{2}}{r^{9-2i}}$ \\ with $r(0)=1$ } \\
		\midrule[0.5pt]
		5 & \tabincell{l}{This \\ article}	& \tabincell{l}{LBS gain: \\ \eqref{ut3.3} with $\mu _{m}=e^3$, $m=1,2,\ldots$} \\
		\midrule[0.5pt]
		6 & \tabincell{l}{This \\ article}	& \tabincell{l}{LBS gain: \\ \tabincell{l}{\eqref{ut3.3} with $\mu _{1}=e^{8}$  and\\ $\mu _{m}=e^{3}$, $m=2,3,\ldots$}} \\	
		\bottomrule[1pt]
	\end{tabular}}
\end{table}

Let $\theta_R =0.1$. To ensure the fairness of the comparison, for Cases 1-6, the control parameters are set as $\left[ a_{1},a_{2},a_{3}\right] ^{\mathrm{T}}=\left[ 3,3,3\right] ^{\mathrm{T}}$ and $\left[ b_{1},b_{2},b_{3}\right]^{\mathrm{T}}=\left[ 0.3,0.8,1.2\right] ^{\mathrm{T}}$, and the initial values are selected as $[ x_{1}(0),$ $x_{2}(0),x_{3}(0)] ^{\mathrm{T}}=\left[ 2,-2,2\right] ^{\mathrm{T}}$ and $\left[ \hat{x}_{1}(0),\hat{x}_{2}(0),\hat{x}_{3}(0)\right] ^{\mathrm{T}}=\left[ 0,0,0\right] ^{\mathrm{T}}$. After calculation, we obtain that $\lambda _{\min }(P)=3.0419$ and $\lambda _{\max }(P)=99.6799$. For Cases 5 and 6, the positive nondecreasing function $\phi (\omega _{m})$ is chosen as $\phi (\omega _{m})=1$, the positive sequence $\bar{\sigma}$ is selected as $\bar{\sigma}_{m}=m-0.9$, $m=1,\ldots ,$ the positive sequence $\underline{\sigma }$ is selected as $\underline{\sigma }_{m}=e^{-m}$, $m=1,\ldots $, and $r_{0}=1$.

Under those controllers, the simulation results are displayed in Figs. \ref{Fig-E2-1}-\ref{Fig-E2-6-2}. It is shown that the proposed control approach performs well with faster convergence speed and better transient performance. For example, the convergence time and peak value of state $x_1$ are significantly decreased from Case 1 to Case 6.

In summary, the above examples and comparisons firmly verify the effectiveness and advantages of the LBS gain approach.

\end{example}

\section{Conclusion}  \label{Conclusion}

This article has studied the global regulation problem for feedforward nonlinear systems with unknown input-output-dependent growth rates. A novel LBS gain approach has been proposed to counteract system uncertainties and nonlinearities. Moreover, a \emph{tanh}-type speed-regulation function is embedded in the switching mechanism to improve control performance. Under the proposed controller, the global regulation of the systems has been achieved with faster convergence speed and better transient performance.

\bibliographystyle{plain}

\end{document}